\documentclass[aps,manuscript,showpacs,showkeys,superscriptaddress]{revtex4}
\usepackage{graphicx}
\usepackage{epsfig}  
\usepackage{epsf}    
\usepackage{dcolumn}
\usepackage{bm}
\usepackage{dcolumn}
\usepackage{textcomp}
\usepackage[tbtags]{amsmath}
\usepackage{amsfonts}
\usepackage{float}
\usepackage{subfig}
\usepackage[]{hyperref}
  \hypersetup{
  unicode=false,          
  pdftoolbar=true,        
  pdfmenubar=true,        
  pdffitwindow=true,     
  pdfstartview={FitH},    
  pdfsubject={Getting the best out of T2K and NOvA},   
  pdfnewwindow=true,      
  pdfcreator={RevTeX},
  colorlinks=true,       
  linkcolor=red,          
  citecolor=blue,        
  urlcolor=blue,           
  }
\usepackage{hypcap}

\def\anu{{\bar\nu}}

\newcommand{\beq}{\begin{equation}}
\newcommand{\eeq}{\end{equation}}
\newcommand{\beqa}{\begin{eqnarray}}
\newcommand{\eeqa}{\end{eqnarray}}

\newcommand{\tx}{{\theta_{12}}}
\newcommand{\ty}{{\theta_{13}}}
\newcommand{\tz}{{\theta_{23}}}

\newcommand{\dl}{{\Delta_{31}}}
\newcommand{\ds}{{\Delta_{21}}}

\newcommand{\atil}{\hat{A}}
\newcommand{\dtil}{\hat{\Delta}}

\newcommand{\dcp}{\delta_{\mathrm{CP}}}
\newcommand{\nova}{NO$\nu$A~}
\newcommand{\pmue}{P(\nu_\mu \rightarrow \nu_e)}

\newcommand{\pmuebar}{P(\bar{\nu}_{\mu} \rightarrow \bar{\nu}_e)}

\newcommand{\dchsq}{\Delta\chi^2}

\begin{document}


\title{Hierarchy sensitivity of \nova in light of T2K $\nu_e$ appearance data}


\author{Suman Bharti}
\email[Email Address: ]{sbharti@phy.iitb.ac.in}
\affiliation{Department of Physics, Indian Institute of Technology Bombay,
Mumbai 400076, India}
\author{Suprabh Prakash}
\email[Email Address: ]{prakash3@mail.sysu.edu.cn}
\affiliation{Harish-Chandra Research Institute, Chhatnag Road, 
Jhunsi, Allahabad 211019, India}
\affiliation{School of Physics and Engineering, Sun Yat-Sen University,
Guangzhou 510275, P. R. China}
\author{Ushak Rahaman \footnote{Corresponding author}}
\email[Email Address: ]{ushak@phy.iitb.ac.in}
\affiliation{Department of Physics, Indian Institute of Technology Bombay,
Mumbai 400076, India}
\author{S. Uma Sankar}
\email[Email Address: ]{uma@phy.iitb.ac.in}
\affiliation{Department of Physics, Indian Institute of Technology Bombay,
Mumbai 400076, India}
\date{\today}
\begin{abstract}
The $\nu_e$ appearance data of T2K experiment has given a glimpse of 
 the allowed parameters in the hierarchy-$\dcp$ parameter space. In this
 paper, we explore how this data affects our expectations regarding the
 hierarchy sensitivity of the \nova experiment. For the favourable combinations
 of hierarchy and $\dcp$, the hierarchy sensitivity of \nova is unaffected
 by the addition of T2K data. For the unfavourable combinations, \nova data
 gives degenerate solutions. Among these degenerate solutions, T2K data prefers 
 IH and $\dcp$ in the lower half plane over NH and $\dcp$ in the upper half plane.
 Hence, addition of the T2K data to NO$\nu$A creates a bias towards IH and 
 $\dcp$ in the lower half plane irrespective of what the true combination
 is.
\end{abstract}
\maketitle

\section{Introduction}
Discovery of neutrino oscillations has led to an explosion of interest in
understanding the fundamental properties of neutrinos. With the data from
the solar and atmospheric neutrino experiments, we have a picture of three
neutrino flavours, $\nu_e$, $\nu_\mu$ and $\nu_\tau$, mixing with one another
to form three light neutrino mass eigenstates $\nu_1$, $\nu_2$ and $\nu_3$.
Measurement of the survival probability of electron neutrinos in the solar 
neutrino experiments \cite{Bahcall:2004ut,Ahmad:2002jz} and that 
of electron anti-neutrinos in KamLAND \cite{Araki:2004mb,Abe:2008aa} led to 
a precise determination of $\Delta_{21} = m_2^2 - m_1^2$ and $\theta_{12}$. 
Measurement of the muon neutrino survival probability by the MINOS 
\cite{Nichol:2012} and T2K \cite{T2Kdisapp} experiments led to the precise
determination of $\sin^2 2 \theta_{23}$ and $|\Delta m^2_{\mu \mu}|$. The data 
indicates that the two mixing angles $\theta_{12}$ and $\theta_{23}$ are quite
large (in fact, $\theta_{23}$ is close to maximal) \cite{Gonzalez-Garcia:2014bfa}
 and $\Delta_{21} \ll |\Delta m^2_{\mu \mu}|$. 
The values of $\Delta_{31} = m_3^2 - m_1^2$ and $\Delta_{32} = m_3^2 - m_2^2$
can be obtained from the relation \cite{Nunokawa:2005nx}
\begin{equation}
 \Delta m^2_{\mu \mu}= \sin^2\tx \dl + \cos^2 \tx \Delta_{32} + \cos \dcp \sin 2\tx \sin \ty \tan \tx \ds.
 \label{dmumu}
\end{equation}

At present only the magnitude of $\Delta m^2_{\mu \mu}$ is known but not its
sign. Since $\Delta_{21} \ll |\Delta m^2_{\mu \mu}|$, the signs of $\Delta_{31}$ and
$\Delta_{32}$ are the same as that of $\Delta m^2_{\mu \mu}$. If $\Delta_{31}$ is 
positive, a likely neutrino mass pattern is $m_3 \gg m_2 > m_1$, which is called
normal hierarchy (NH). If $\Delta_{31}$ is negative, the neutrino mass pattern is
likely to be $m_2 > m_1 \gg m_3$, which is called inverted hierarchy (IH). It is
of course possible to have $\Delta_{31}$ positive or negative when all the three
neutrino masses are quasi-degenerate. In such a situation also, positive $\Delta_{31}$ 
is called NH and negative $\Delta_{31}$ is called IH. 

In the past few years, reactor neutrino experiments DoubleCHOOZ, 
Daya Bay and RENO \cite{An:2012eh,Ahn:2012nd,Abe:2012tg}, with baslines $\sim 1$ km, 
have measured $\theta_{13}$ to be non-zero. The moderately large value of $\theta_{13}$
has given hope that the outstanding questions related to neutrino oscillations
can soon be answered. These questions are
\begin{itemize}
\item
What is correct neutrino mass hierarchy, NH or IH? 
\item 
What is the true octant of $\theta_{23}$? Is $\theta_{23} < \pi/4$
or $> \pi/4$?
\item
Is there CP violation in the neutrino sector? If yes, what is the 
value of the CP violating phase $\dcp$?
\end{itemize}
All these questions can be answered by the measurement of the oscillation 
probabilities $P(\nu_\mu \to \nu_e)$ and $P(\bar{\nu}_\mu \to \bar{\nu}_e)$ 
at the long baseline neutrino experiments T2K and NO$\nu$A. 
T2K experiment has already taken significant amount of
data and \nova experiment has begun its run. In this paper, we address the
question: How does the data of T2K modify our expectations regarding the 
mass hierarchy determination capability of NO$\nu$A?

\section{Hierarchy-$\dcp$ degeneracy}

The oscillation probabilities $P(\nu_\mu \to \nu_e)$ and $P(\bar{\nu}_\mu \to \bar{\nu}_e)$
can be calculated in terms of the three mixing angles 
$\theta_{12}$, $\theta_{13}$ and $\theta_{23}$, the mass-squared
differences $\Delta_{21}$ and $\Delta_{31}$ and the CP violating phase $\dcp$.
In long baseline experiments, however, the neutrinos travel long distances through
earth matter and undergo coherent forward scattering. The effect of this scattering
is taken into account through the Wolfenstein matter term \cite{msw1}
\begin{equation}
A \ ({\rm in \ eV^2}) = 0.76 \times 10^{-4} \rho \ ({\rm in \ gm/cc}) \ E \ ({\rm in \ GeV}),
\end{equation}
where $E$ is the energy of the neutrino and $\rho$ is density of the matter.
The interference between $A$ and $\Delta_{31}$ leads to the modification of neutrino
oscillation probability due to matter effects. The expression for $P(\nu_\mu \to \nu_e)$
is given by \cite{Cervera:2000kp,Freund:2001pn} 
\begin{eqnarray}
\pmue & = & 
\sin^2 2 \ty \sin^2 \tz\frac{\sin^2\dtil(1-\atil)}{(1-\atil)^2} \nonumber\\ 
& & +\alpha \cos \ty \sin2\tx \sin 2\ty \sin 2\tz \cos(\dtil+\dcp)
\frac{\sin\dtil \atil}{\atil} \frac{\sin \dtil(1-\atil)}{1-\atil} \nonumber\\
 & & +\alpha^2 \sin^2 2 \tx \cos^2 \ty \cos^2 \tz 
 \frac{\sin^2 \dtil \atil}{\atil^2}.
\label{pmue-exp}
\end{eqnarray}
where $\hat{\Delta} = \Delta_{31}L/4E$, $\hat{A} = A/\Delta_{31}$ and 
$\alpha = \Delta_{21}/\Delta_{31}$. For anti-neutrinos, 
$P(\bar{\nu}_\mu \to \bar{\nu}_e)$ is given by a similar expression
with $\dcp \to - \dcp$ and $A \to -A$.

\begin{figure}[t]
\centering
\includegraphics[width=0.70\textwidth]{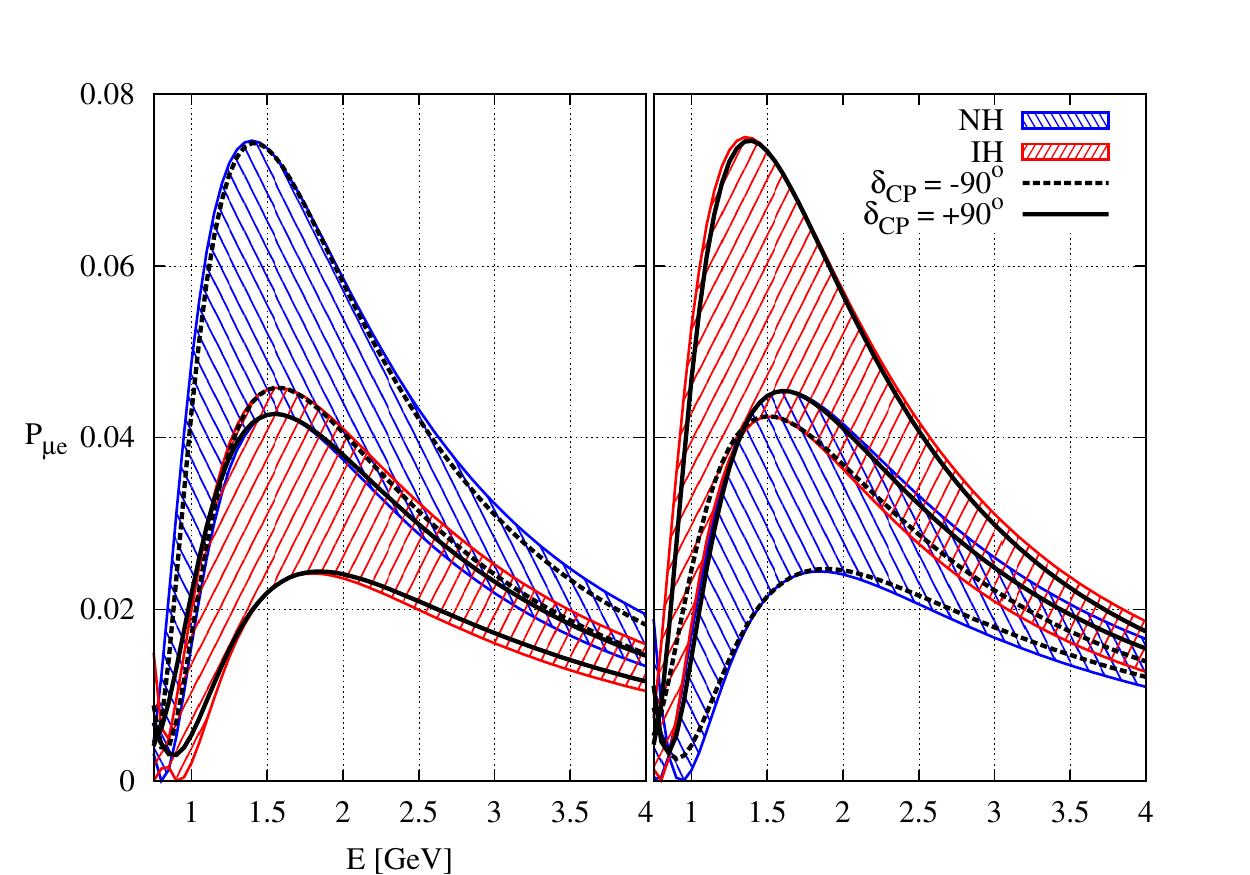}
\caption{$\pmue$ (left panel) and $\pmuebar$ (right panel) vs. energy for
NO$\nu$A. Variation of $\dcp$ 
leads to the blue (red) bands for NH (IH). 
The plots are drawn for
maximal $\tz$ and other neutrino parameters given in the text.}
\label{prob}
\end{figure}

$P(\nu_\mu \to \nu_e)$ is sensitive to the neutrino mass hierarchy because 
both $\hat{\Delta}$ and $\hat{A}$ change sign under a change of sign of $\Delta_{31}$.
The term $\sin[(1-\hat{A})\hat{\Delta}]/(1 - \hat{A})$ undergoes a change 
under the sign change of $\hat{A}$. This change may or may not be measurable because
value of $\dcp$ is completely unknown at the moment. For certain choices of hierarchy 
and values of $\dcp$, the change in the first term of eq.~(\ref{pmue-exp}) arising due to changing the hierarchy
can be compensated by a change in the second term caused by choosing a wrong value of $\dcp$.
It was shown that \nova experiment \cite{nova_tdr} can determine the neutrino mass hierarchy, that is 
measure the change induced by the matter term, for the following two favourable cases:
\begin{itemize}
\item hierarchy is NH and $\dcp$ is in the lower half plane $(-180^\circ \leq \dcp \leq 0)$ and
\item hierarchy is IH and $\dcp$ is in the upper half plane $(0 \leq \dcp \leq 180^\circ)$.  
\end{itemize}
If nature had chosen either of these favourable cases, \nova can determine 
both the hierarchy and the half plane of $\dcp$.
For the two unfavourable cases,
\begin{itemize}
\item hierarchy is NH and $\dcp$ is in the upper half plane $(0 \leq \dcp \leq 180^\circ)$ and
\item hierarchy is IH and $\dcp$ is in the lower half plane $(-180^\circ \leq \dcp \leq 0)$,  
\end{itemize}
an analysis of \nova data gives degenerate solutions. Hence \nova alone is unable to determine the
hierarchy for all possible combinations of hierarchy amd $\dcp$ \cite{Mena:2004sa,Prakash:2012az}. 
This is illustrated in the plots of $P(\nu_\mu \to \nu_e)$ and $P(\bar{\nu}_\mu
\to \bar{\nu}_e)$ for NO$\nu$A, shown in fig.~\ref{prob}.  In this paper, we study how the presently 
collected neutrino data from the T2K \cite{T2Kapp, T2Kdisapp} modifies these expectations from NO$\nu$A.  

\section{Simulation and Calculation Details}
We have mentioned in the previous section that \nova can determine
the hierarchy by itself, for favourable hierarchy-$\dcp$ combinations.
It was suggested that a combination of data from \nova and T2K may be
able to determine the hierarchy for unfavourable combinations also
\cite{Mena:2004sa,Prakash:2012az}. Since T2K has already produced about one year
of neutrino data, we now explore the hierarchy determination capability
of \nova in light of this data. 

A difficulty arises in combining the simulations of \nova with the 
data of T2K. The data of T2K contain random fluctuations but the
simulations of \nova do not. For data without fluctuations, 
$\chi^2_{\rm min}$ is zero whereas for data with fluctuations, 
$\chi^2_{\rm min}$ is expected to be equal to the degrees of freedom.
The question then arises: How to combine the simulations and data 
in such a way that we can generate practical definitions of
$\chi^2_{\rm min}$ and $\Delta \chi^2$ which can be used in analysis?
The only practical approach is to simulate \nova data with fluctuations.
\subsection{Simulation of \nova experiment}

\nova \cite{nova_tdr} is a long baseline neutrino oscillation experiment capable
of measuring the survival probability $P(\nu_\mu \to \nu_\mu)$
and the oscillation probability $P(\nu_\mu \to \nu_e)$. The 
NuMI beam at Fermilab, with the power of 700 kW which corresponds
to $6\times 10^{20}$ protons on target (POT) per year, produces the neutrinos. 
The far detector consists of 14 kton of totally active scintillator 
material and is located 810 km away at a $0.8^\circ$ off-axis location.
Due to the off-axis location, the flux peaks sharply at
2 GeV, which is close to the energy of maximum oscillation of 1.4 GeV.
It has started taking data in 2014 and is expected to run three years in
neutrino mode and three years in anti-neutrino mode. In our simulations,
we have taken the retuned signal acceptance and background rejection 
factors from \cite{Kyoto2012nova,Agarwalla:2012bv}.

In doing the simulations, we have used the "true" values of the 
neutrino parameters to be their central values, namely $\sin^2 
\theta_{12} = 0.3$, $\sin^2 2 \theta_{13} = 0.084$, $\sin^2 \theta_{23}
= 0.514$, $\Delta_{21} = 7.5 \times 10^{-5}$ eV$^2$ and 
$\Delta m^2_{\mu \mu} = \pm 2.4 \times 10^{-3}$ eV$^2$ \cite{Gonzalez-Garcia:2014bfa, T2Kdisapp}.   
The values for $\Delta_{31}$(NH) and $\Delta_{31}$(IH) were derived
from $\Delta m^2_{\rm eff}$ using the expression given in eq.~(\ref{dmumu}).
Simulations were done with NH as the true hierarchy as well as with IH.
The following true values of $\dcp$ were chosen as inputs in the
simulations: $-135^\circ$, $-90^\circ$, 
$-45^\circ$, $0$, $45^\circ$, $90^\circ$, $135^\circ$ and $180^\circ$. We used these 
true values as inputs in the software GLoBES \cite{Huber:2004ka,Huber:2007ji}
to calculate the expected $\nu_e$ appearance events in $i$th energy 
bin $N_{i}^{\rm exp}$.

To take into account the possible fluctuations in the expected data,
we took $N_i^{\rm exp}$, 
and gave it as an input to the Poissonian random number generator
code \cite{random}. This code generated 100 Poissonian random numbers
whose mean is $N_i^{\rm exp}$. We repeated this procedure for all 
the energy bins. Thus, we generated 100 possible event numbers for
each bin. We collected the first of the 100 numbers from $i$th energy bin 
and labelled it $N_i^{\rm data\#1}$. By collecting the second of the 
100 numbers from the $i$th energy bin we obtain $N_i^{\rm data\#2}$ etc.
Thus, we obtain 100 independent simulations of the $\nu_e$ appearance 
data which include the random Poissonian fluctuations expected in
counting experiments.

The ``theoretical" event rates, corresponding to this data, are calculated 
for various test values of the neutrino parameters. 
The test values for 
$\sin^2 2\theta_{13}$ ($\sigma (\sin^2 2 \theta_{13}) = 5\%$) \cite{Gonzalez-Garcia:2014bfa} and 
$\Delta m^2_{\mu\mu}$ ($\sigma(\Delta m^2_{\mu \mu})=3\%$) \cite{Itow:2001ee}
are selected within the $\pm 2 \sigma$ range of the central values. 
Since $\sin^2 \theta_{23}$ is not-so well constrained, its test 
values are picked within the $\pm3\sigma$ range: [0.35, 0.65]. 
Test values of $\dcp$ spanned its total allowed range: [$-180^\circ$, $180^\circ$].
With the selected test values as inputs to GLoBES, we calculated
$N_i^{\rm test}$ for $\nu_e$ appearance as functions of the 
test values of neutrino parameters. As before, here $i$ stands for the $i$th energy bin. 

We compute the Poissonian $\chi^2$ between $N_i^{\rm data\#1}$ and 
$N_i^{\rm test}$ using the formula \cite{Coloma:2012ji}
\begin{eqnarray}
\chi^2(1) &=& \sum_i 2[{(N_i^{\rm test} - N_i^{\rm data\#1}) + N_i^{\rm data\#1}
\times \ln(N_i^{\rm data\#1}/N_i^{\rm test})}] + \sum_j [2\times N_j^{\rm test}] \nonumber \\
& & + \chi^2({\rm prior})
\label{poisionian}
\end{eqnarray}
where $i$ stands for bins for which $N_i^{\rm data\#1}\neq 0$ and 
$j$ stands for bins for which $N_j^{\rm data\#1} = 0$.
$\chi^2({\rm prior})$
is the prior added due to the deviation of the test values of neutrino
parameters from their best fit values. It is defined by 
\begin{eqnarray}
\chi^2({\rm prior}) &=&  ((\sin^2 2\ty ({\rm test}) - 0.084)/(0.05\times0.084))^2 + \nonumber\\  
 && ((\sin^2 2\tz ({\rm test})-4\times0.514\times0.486)/(0.02\times4\times0.514\times0.486))^2 + \nonumber\\
&& ((|\Delta m^2_{\mu \mu} ({\rm test})|-2.40\times 10^{-3})/(0.03\times2.40\times 10^{-3}))
\end{eqnarray}
Since $N_i^{\rm test}$ is a function of the test values of the 
neutrino paramters, $\chi^2(1)$ is also a function of the same 
test values. We find the minimum value of $\chi^2(1)$ and subtract
it from each of the values of $\chi^2(1)$ to obtain $\dchsq(1)$.
It is zero for those test values of
neutrino parameters for which $\chi^2(1)$ is minimum.  
Since $N_i^{\rm data\#1}$ contains fluctuations, the test values of
neutrino parameters for which $\dchsq(1)$ vanishes are not the same as
the input values used in the simulations.  

Next we marginalize $\dchsq(1)$ over
$\sin^2 2\theta_{13}$, $\sin^2 \theta_{23}$ and $|\Delta m^2_{\rm eff}|$
but not over $\dcp$ and hierarchy and label the result $\dchsq_m(1)$.
Thus $\dchsq_m(1)$ is a function 
of test $\dcp$ and test hierarchy. As mentioned above, $\dchsq_m(1)$ is zero for 
some value of test $\dcp$ and test hierarchy. We then compute 
$\dchsq_m(2)$ from $N_i^{\rm data\#2}$ using the procedure described above.
$\dchsq_m(2)$ also vanishes for some value of test $\dcp$ and test 
hierarchy but these values need not be the same ones for which 
$\dchsq_m(1)$ vanishes. Treating $N_i^{\rm data\#p}$ $(1 \leq p \leq 100)$
as the ``data", we compute 100 different sets of $\dchsq_m (p)$  
as functions of test $\dcp$ and test hierarchy. Each of these sets contains
a zero element at some test $\dcp$ and test hierarchy. However, for a given hierarchy and a 
given test value of $\dcp$, a large number of $\dchsq_m(p)$ will be non-zero. 
We take the average of these 100 $\dchsq_m$ values to
finally obtain $\overline{\dchsq_m}$ as a function of the test values of $\dcp$ and hierarchy. 
The quantity $\overline{\dchsq_m}$ is equivalent to the  $\dchsq$ obtained in 
simulations where the ``data" was simulated without fluctuations.

\begin{figure}[H]
\centering
\includegraphics[width=0.8\textwidth]{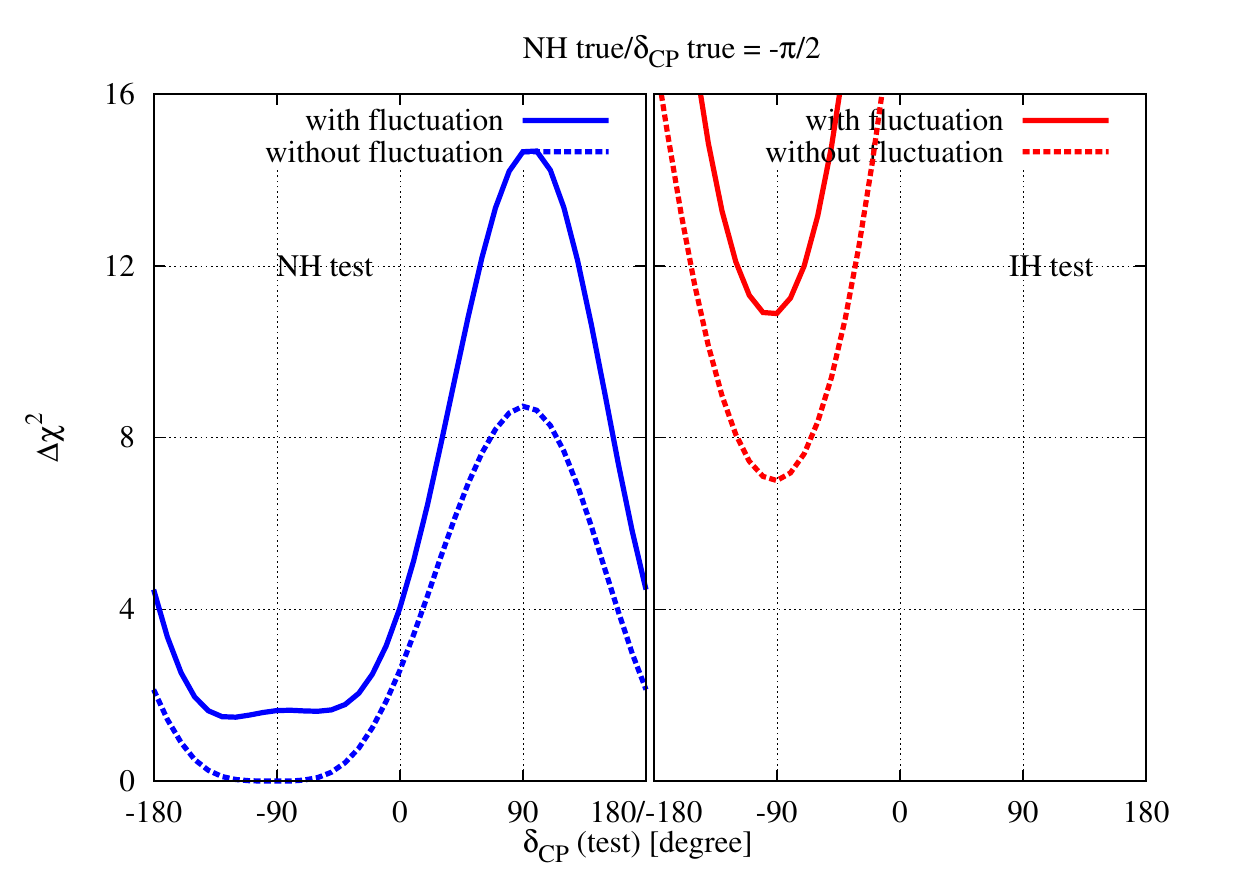}
\caption{\footnotesize{Comparison between $\dchsq$ vs test values of $\dcp$ from \nova 
simulation without fluctuations and with fluctuations. NH is the true hierarchy and 
true value of $\dcp$ is $-90^\circ$.}}
\label{NH-90g}
\end{figure}

\begin{figure}[H]
\centering
\includegraphics[width=0.8\textwidth]{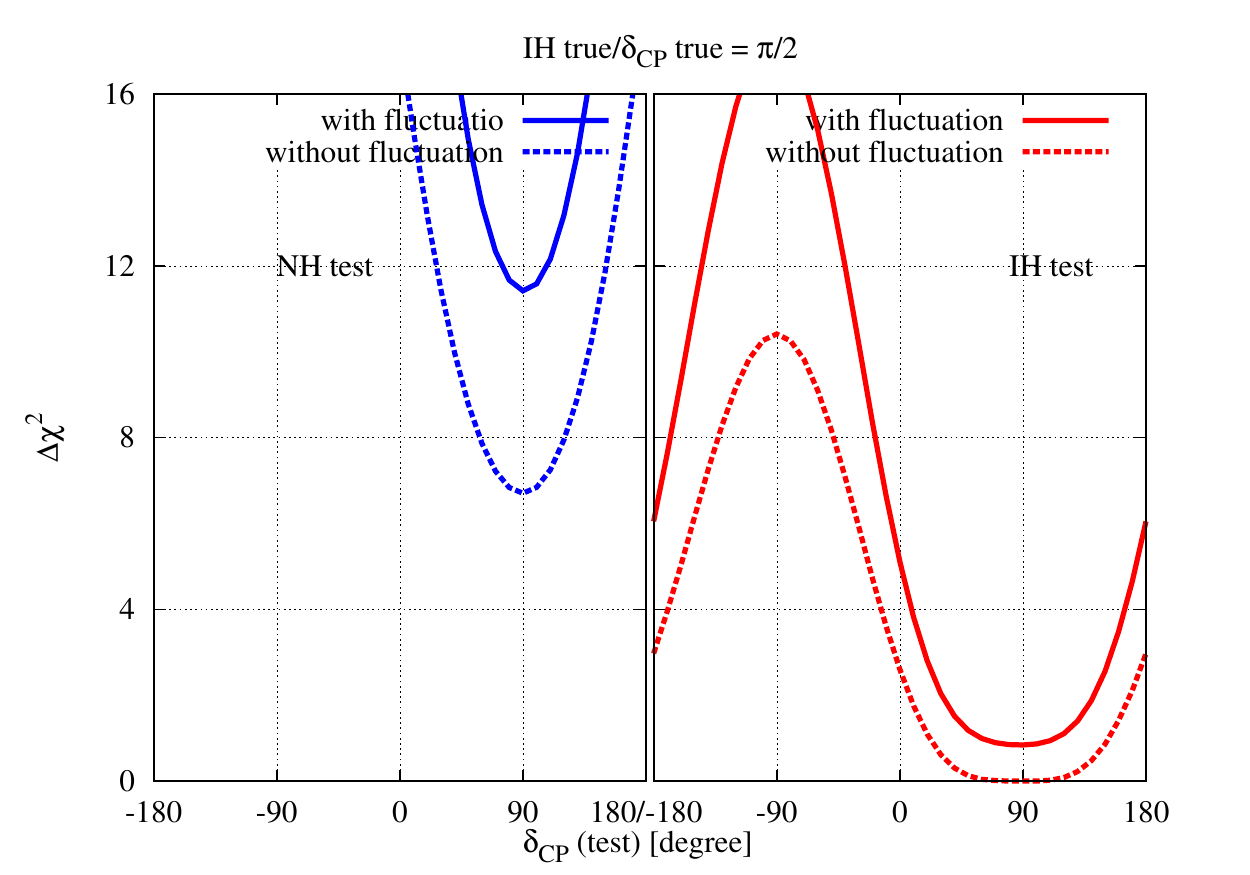}
\caption{\footnotesize{Comparison between $\dchsq$ vs test values of $\dcp$ from \nova 
simulation without fluctuations and with fluctuations. IH is the true hierarchy and 
true value of $\dcp$ is $90^\circ$.}}
\label{IH90g}
\end{figure}

\begin{figure}[H]
\centering
\includegraphics[width=0.8\textwidth]{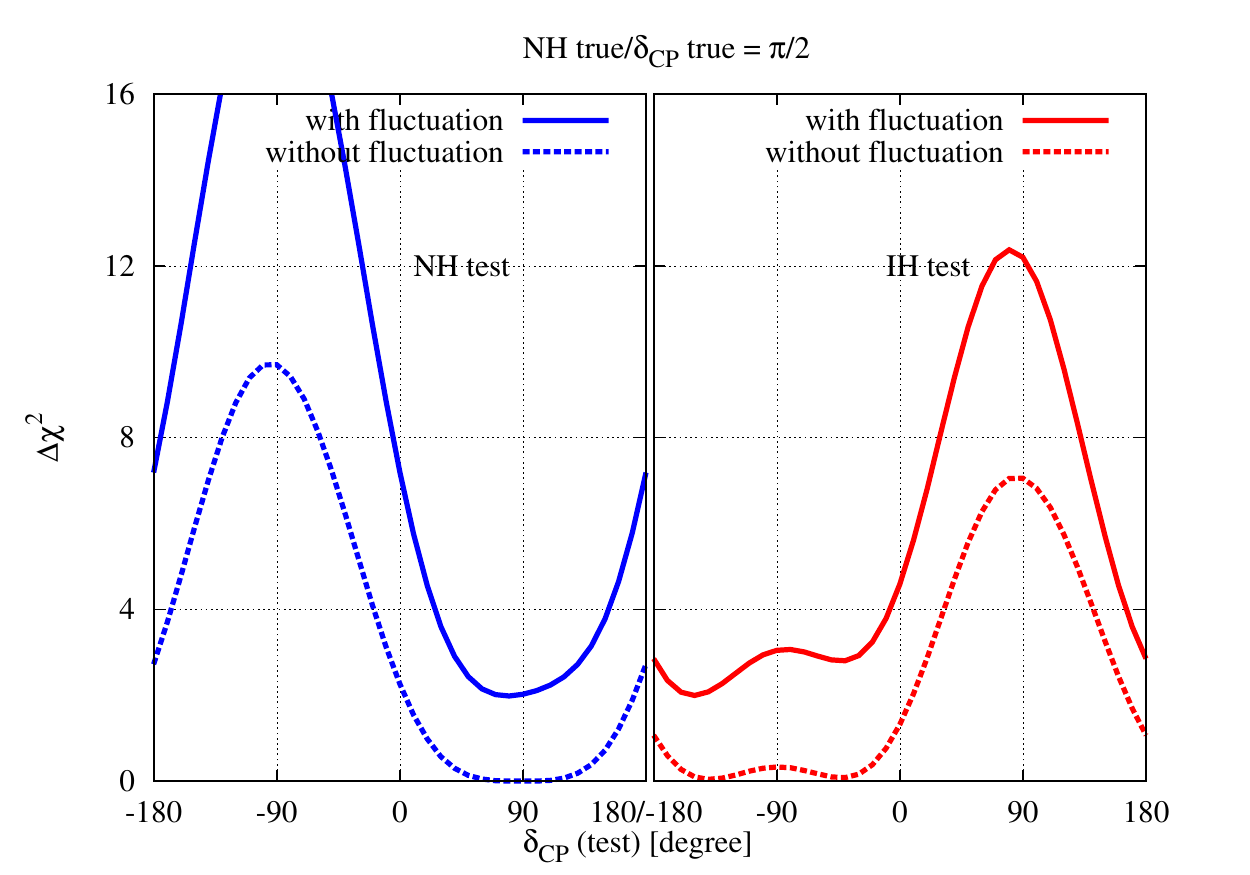}
\caption{\footnotesize{Comparison between $\dchsq$ vs test values of $\dcp$ from \nova 
simulation without fluctuations and with fluctuations. NH is the true hierarchy and 
true value of $\dcp$ is $90^\circ$.}}
\label{NH90g}
\end{figure}

\begin{figure}[H]
\centering
\includegraphics[width=0.8\textwidth]{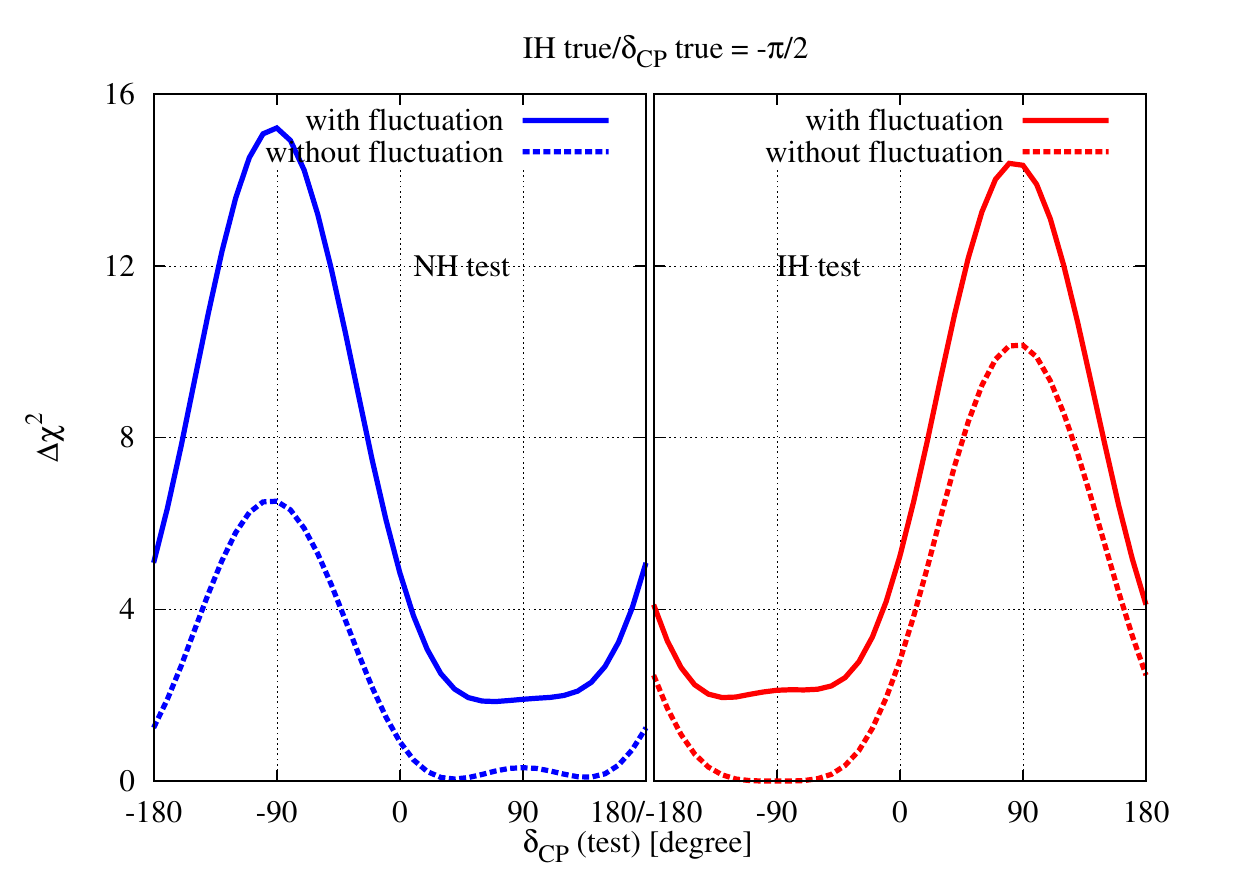}
\caption{\footnotesize{Comparison between $\dchsq$ vs test values of $\dcp$ from \nova 
simulation without fluctuations and with fluctuations. IH is the true hierarchy and 
true value of $\dcp$ is $-90^\circ$.}}
\label{IH-90g}
\end{figure}
In figures \ref{NH-90g} to \ref{IH-90g}, we have plotted the hierarchy-discriminating   
$\dchsq$ vs test $\dcp$. All these figures contain two curves:
One curve is obtained by our procedure of calculating
$\overline{\dchsq_m}$ from \nova simulation with fluctuations and
the other curve is obtained by doing simulations without fluctuations. 
The plots show hierarchy discrimination for the 
for two most favourable hierarchy - $\dcp$ combinations (NH and $\dcp=-90^\circ$ 
in fig.~2 
$\&$ IH and $\dcp=90^\circ$ in fig.~3) and two most unfavourable hierarchy - $\dcp$ 
combinations (NH and $\dcp=90^\circ$ in fig.~4 $\&$ IH and $\dcp=-90^\circ$ in fig.~5). 
We see that for these four cases, $\overline{\dchsq_m}$ matches qualitatively with $\dchsq$.
$\overline{\dchsq_m}$ never vanishes because of the averaged effect of the fluctuations but the
physics remains same in simulations both with and without fluctuations. 
This verifies our earlier statement that $\overline{\dchsq_m}$ correctly represents 
the hierarchy sensitivity.

\begin{figure}[H]
\centering
\includegraphics[width=0.8\textwidth]{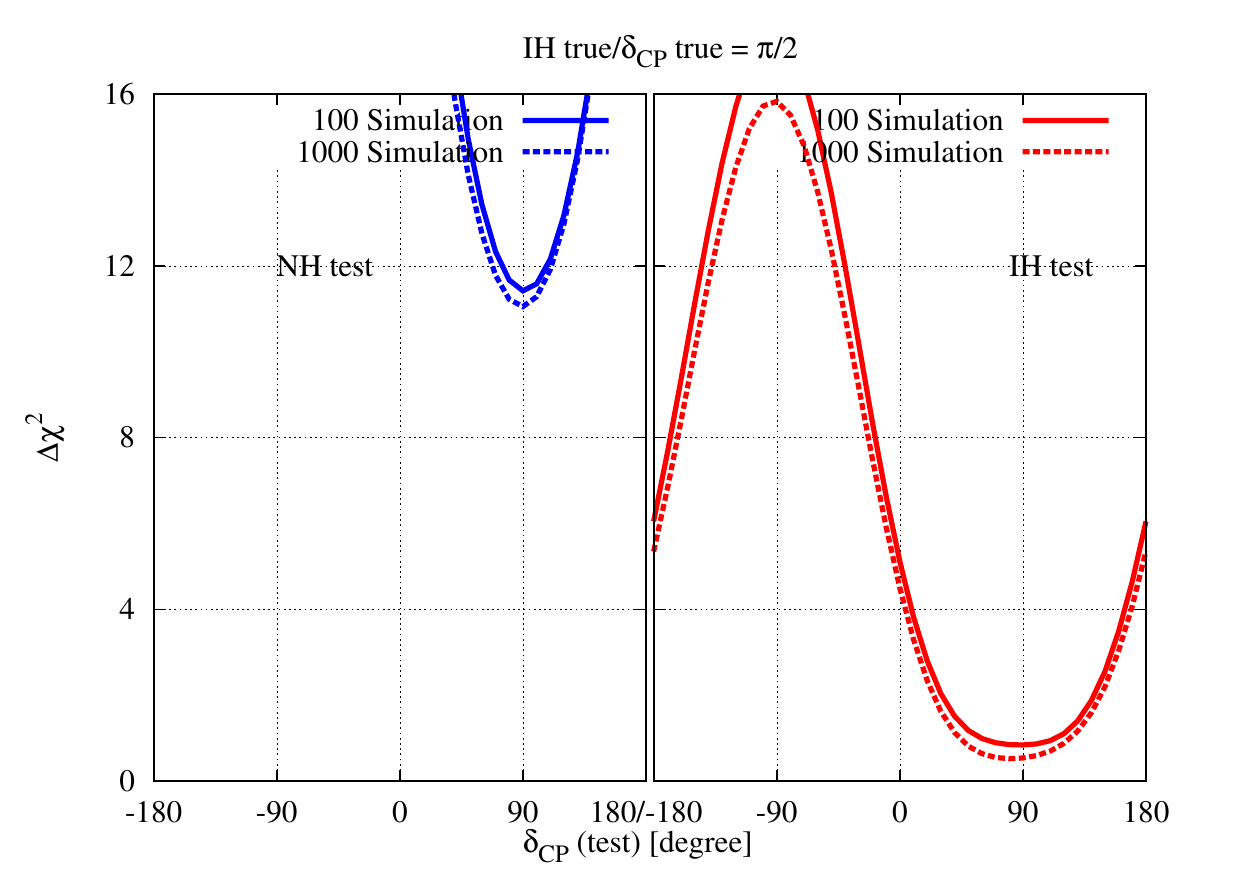}
\caption{\footnotesize{Comparison between $\dchsq$ vs test values of $\dcp$ from \nova 
simulation with fluctuations for 100 independent simulations and 1000 independent simulations. IH is the true hierarchy and 
true value of $\dcp$ is $90^\circ$.}}
\label{IH90-100vs1000}
\end{figure}

To check the stability of this averaging method, 
we have also done 1000 independent simulations of NO$\nu$A. That is, 
we have generated 1000 random Poissonian event numbers for each bin, whose 
mean is equal to the event number of that bin.  
Then we followed the above procedure to calculate $\overline{\dchsq_m}$. 
In figure \ref{IH90-100vs1000}, we have compared the $\overline{\dchsq_m}$ from 100 
independent simulations with that of 1000 independent simulations for IH and $\dcp=90^\circ$. 
We see that the $\overline{\dchsq_m}$s from both the simulations match quite closely. 
This holds true for other hierarchy-$\dcp$ combinations as well. 
Thus the values of $\overline{\dchsq_m}$s, derived by our simulation, 
are stable and we will use this method of 100 independent 
simulations to determine the hierarchy sensitivity of 
\nova after adding present T2K data.
\subsection{T2K Calculation}

T2K is a long baseline neutrino oscillation experiment with the
$\nu_\mu$ beam from the J-PARC accelerator in Tokai to the 
Super-Kamiokande detector 295 km away. The accelerator is oriented
such that the detector is at 2.5$^\circ$ off-axis location. 
Super-Kamiokande is a 22.5 kton fiducial mass water Cerenkov detector, capable of
good discrimination between electron and muon neutrino interactions \cite{Itow:2001ee}. 
The neutrino flux peaks sharply at 0.7 GeV which is also the 
energy of the first oscillation maximum. T2K experiment started taking
data in 2009 and ran in neutrino mode with $6.6 \times 10^{20}$ 
POT till 2013 \cite{T2Kapp,T2Kdisapp}. Presently they are taking data in
anti-neutrino mode.

The $\nu_e$ appearance data of T2K were published and analyzed in
ref. \cite{T2Kapp}. They find the best fit point to be normal
hierarchy with $\dcp = -90^\circ$. In general both hierarchies with
the $\dcp$ values in the lower half plane are allowed at 2 $\sigma$,
whereas $\dcp$ values in the upper half place are disfavoured for
both hierarchies. 

\begin{figure}[H]
\centering
\includegraphics[width=0.8\textwidth]{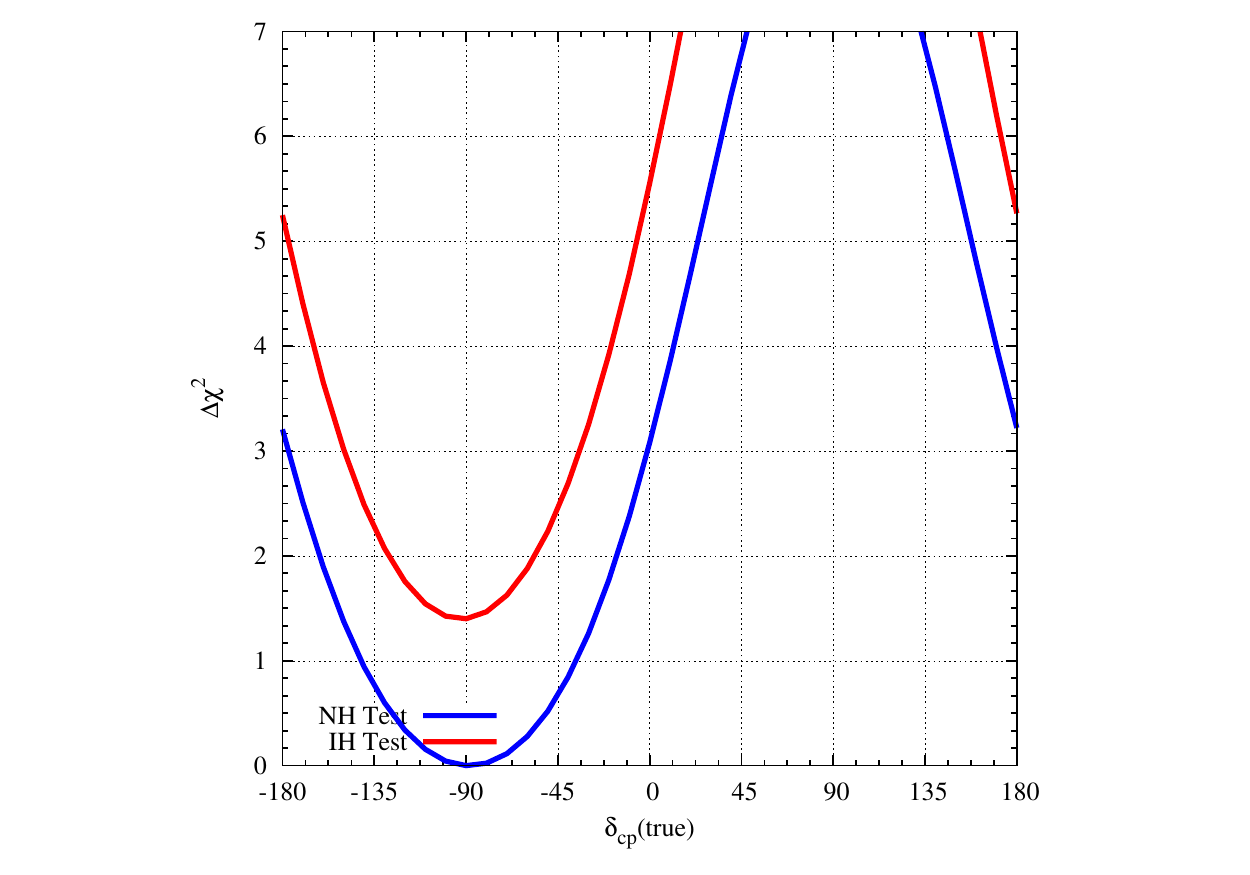}
\caption{\footnotesize{$\dchsq$ vs test $\dcp$ plot for T2K neutrino appearance data}}
\label{T2K}
\end{figure}

From fig. (4) of \cite{T2Kapp}, we get the binned event rates $N_{i}^{\rm data}$
as a function of reconstructed neutrino energy for electron appearance. 
Using GLoBES software, we calculated the electron appearance events $N_{i}^{\rm test}$
for the energy bin $i$ and as a function of the neutrino test parameters 
$|\Delta m^2_{\mu\mu}|$, $\sin^2 2 \theta_{13}$, $\sin^2 \theta_{23}$, $\dcp$ and hierarchy. 
Then we calculated Poissonian $\chi^2$ as a function of 
the test parameters using the formula given in eq. \ref{poisionian}.
The minimum of the $\chi^2$ is obtained and is subtracted from all 
values of $\chi^2$s to get $\dchsq$ as a function of test parameters. 
This $\dchsq$ is marginalized over $\Delta m^2_{\mu\mu}$, $\sin^2 \theta_{13}$
and $\sin^2 \theta_{23}$ but not over test $\dcp$ and test hierarchy.
We have plotted this $\dchsq$ in fig. \ref{T2K} as a function of test $\dcp$ 
for test hierarchy NH as well as IH.

\subsection{Combining \nova simulations with T2K data}

In the next step, we would like to explore how the T2K data modifies 
the hierarchy determination capability of NO$\nu$A. As described 
earlier, we have a hundred different sets of $\chi^2(p)$ ($p = 1,2,...,100$) 
each as a function of the test values of neutrino parameters, for the 
100 simulations of NO$\nu$A. We also have $\chi^2$ of T2K as a function
of the same test values. We now define 
\begin{equation}
\chi^2(p)({\rm tot}) = \chi^2(p) + \chi^2({\rm T2K}).
\end{equation}
In the above addition, we have taken care that  
the test values of neutrino parameters are the
same for both $\chi^2(p)$ and $\chi^2({\rm T2K})$.
Note that $\chi^2(p)$ includes the prior coming due to the deviation
of the test values of neutrino parameters from their best fit values.  
From $\chi^2(p)({\rm tot})$ we obtain $\overline{\dchsq_m}({\rm tot})$ using the same
procedure that was used to calculate $\overline{\dchsq_m}$ from $\chi^2 (p)$,
that was described in subsection~3.1. This quantity shows how the hierarchy
determination capability of \nova is modified by the T2K data. To simplify the 
notation a little, we label this quantity as $\dchsq_{\rm HR}$, i.e. the $\dchsq$
for hierarchy resolution. In the 
next section, we discuss our results where we have calculated 
$\dchsq_{\rm HR}$
for various different true hierarchy-$\dcp$ combinations.

\section{Results}
We have calculated $\dchsq_{\rm HR}$ for a number of combinations of
true values of hierarchy and $\dcp$, both favourable and unfavourable.
In this section we give the a series of plots of $\dchsq_{\rm HR}$ as a function of
test $\dcp$ for both of the test hierarchy being the true hierarchy and the
test hierarchy being the wrong hierarchy. If $\dchsq_{\rm HR} \geq 4$ for all
the values of test $\dcp$ when the test hierarchy is the wrong hierarchy, then
the wrong hierarchy can be ruled out at $\geq 95\%$ confidence level. For
the cases where this is not true, the hierarchy determination is not possible.
We present our results in the following progression.

\subsection{NH as the true hierarchy and true $\dcp = -135^\circ,\ -90^\circ,\ -45^\circ$}
\begin{figure}[H]
\centering
\includegraphics[width=1.0\textwidth]{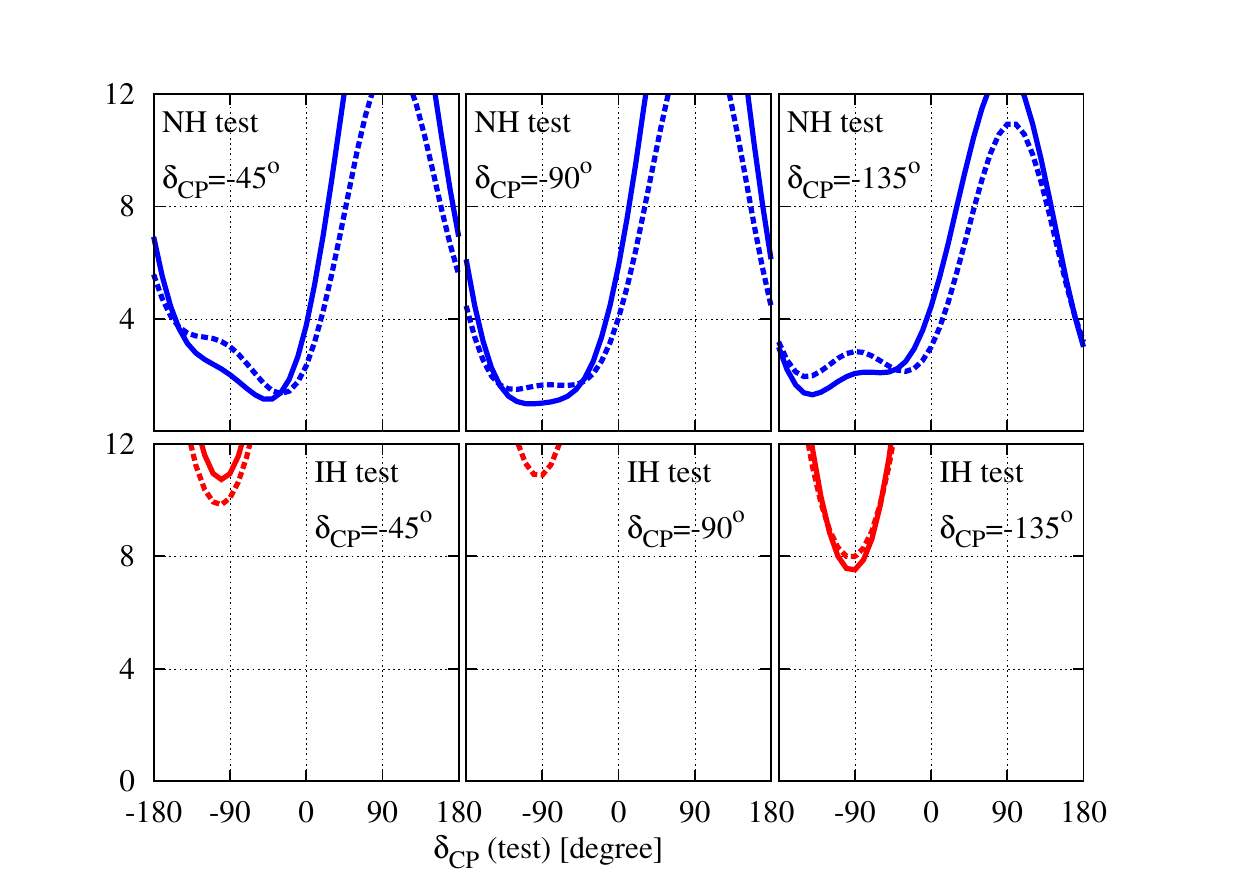}
\caption{\footnotesize{$\dchsq$ vs test $\dcp$ plot for NH true and true $\dcp$ in lower half plane. The upper
(lower) panel shows the plot for test hierarchy NH (IH). True $\dcp$ values 
are written on the panels. The solid (dashed) lines give hierarchy determination capability of \nova
as a function of test values of $\dcp$, with (without) the addition of T2K data.}}
\label{NH-fav} 
\end{figure}

Here all the values of true $\dcp$ are in the lower half plane and hence all the three
cases are favourable for the hierarchy determination by NO$\nu$A. Fig.~\ref{NH-fav} shows the plots
for NH (IH) as the test hierarchy in the upper (lower) panels. As we can see, in all
the lower panels $\dchsq_{\rm HR} \geq 7$, meaning that the wrong hierarchy can be ruled
out quite effectively. We also find, from the lower panels, that the addition of T2K 
data does not lead to any change in the conclusions one obtains from the simulations
of NO$\nu$A. For the upper panels, where the test hierarchy is the true hierarchy, 
the minimum value of $\dchsq_{\rm HR} \simeq 2$ is obtained for value of test $\dcp$
in the same half plane as the input value of true $\dcp$. The non-zero value of 
minimum $\dchsq_{\rm HR}$, as explained in the previous section, arises due to 
taking the average of a hundred simulations. 

\subsection{IH as the true hierarchy and true $\dcp = 45^\circ,\ 90^\circ,\ 135^\circ$}
\begin{figure}[H]
\centering
\includegraphics[width=1.0\textwidth]{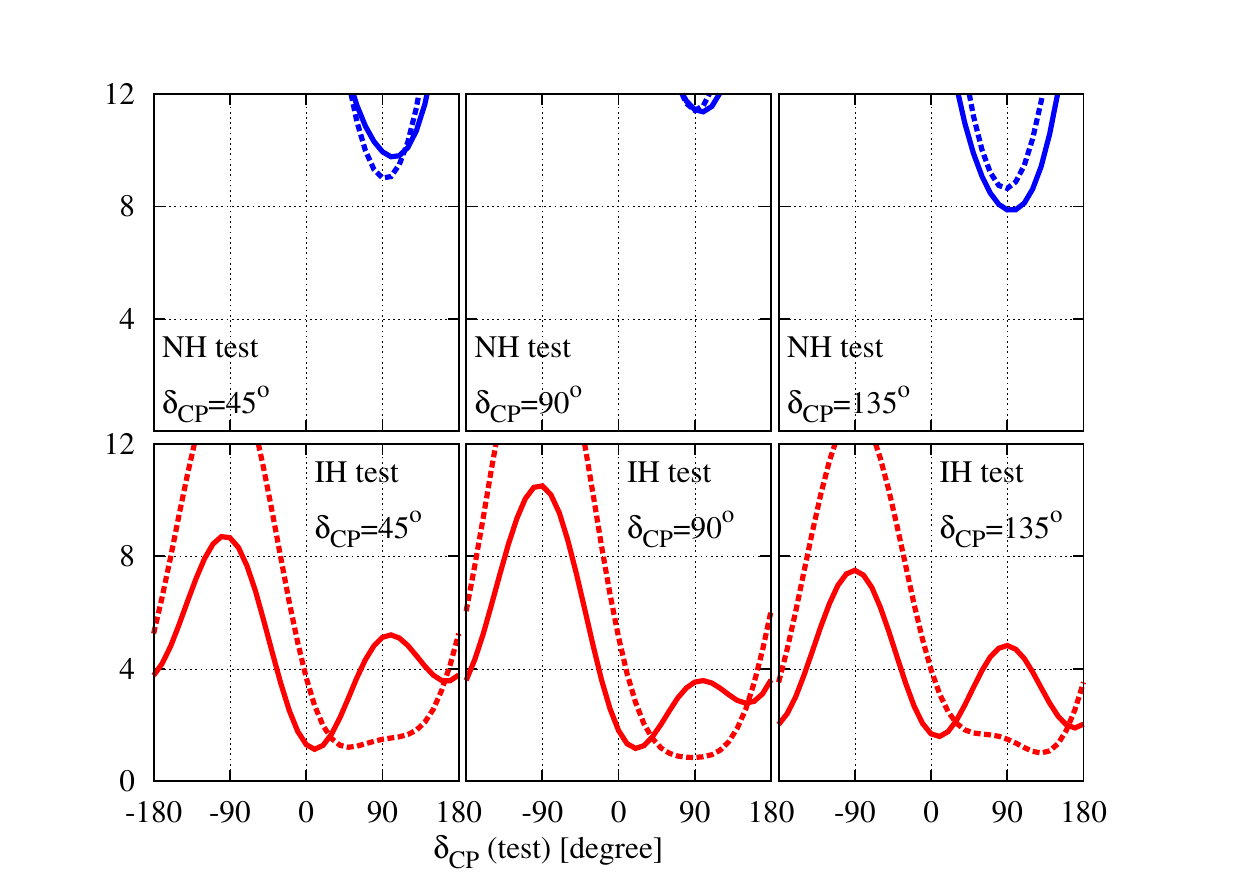}
\caption{\footnotesize{$\dchsq$ vs test $\dcp$ plot for IH true and true $\dcp$ in upper half plane. The upper
(lower) panel shows the plot for test hierarchy NH (IH). True $\dcp$ values 
are written on the panels. The solid (dashed) lines give hierarchy determination capability of \nova
as a function of test values of $\dcp$, with (without) the addition of T2K data.}}
\label{IH-fav} 
\end{figure}

These three cases are also favourable for the hierarchy determination by \nova alone.
Fig.~\ref{IH-fav} shows the plots for NH (IH) as the test hierarchy in the upper (lower) panels.
Here we find that the $\dchsq_{\rm HR} \geq 9$ in all the upper panels which means 
that the wrong hierarchy can be ruled out at nearly $3~\sigma$ level. Looking at the
lower panels, we find a minimum $\dchsq_{\rm HR}$ of about 1 close to test $\dcp
\sim 30^\circ$. This occurs because of the clash between the \nova simulation and
T2K data. T2K data disfavours IH and $\dcp$ in the upper half plane. In fact,
the point IH-$\dcp=90^\circ$ has a $\dchsq = 6$ from the T2K data. However, in our
calculations, we obtain a lower $\dchsq_{\rm HR}$ for test hierarchy IH and test $\dcp$ in the 
upper half plane when \nova simulation is combined with T2K data 
due to the following reason. 
The point favoured by \nova simulation is disfavoured by T2K data and vice verse.
Therefore the combination of the two has a minimum $\dchsq_{\rm HR}$ at some intermediate
point. The reason why the points with IH and test $\dcp$ in upper half plane are not
disfavoured by the combined data is because the hierarchy discrimination capability of the full run
of \nova outweighs the corresponding discrimination of the current T2K neutrino run.
Hence these points, if they happen to be the true points, will be favoured by \nova
(and by \nova plus T2K) even though they are presently disfavoured by T2K. 


\subsection{NH as the true hierarchy and true $\dcp = 45^\circ,\ 90^\circ,\ 135^\circ$}
\begin{figure}[H]
\centering
\includegraphics[width=1.0\textwidth]{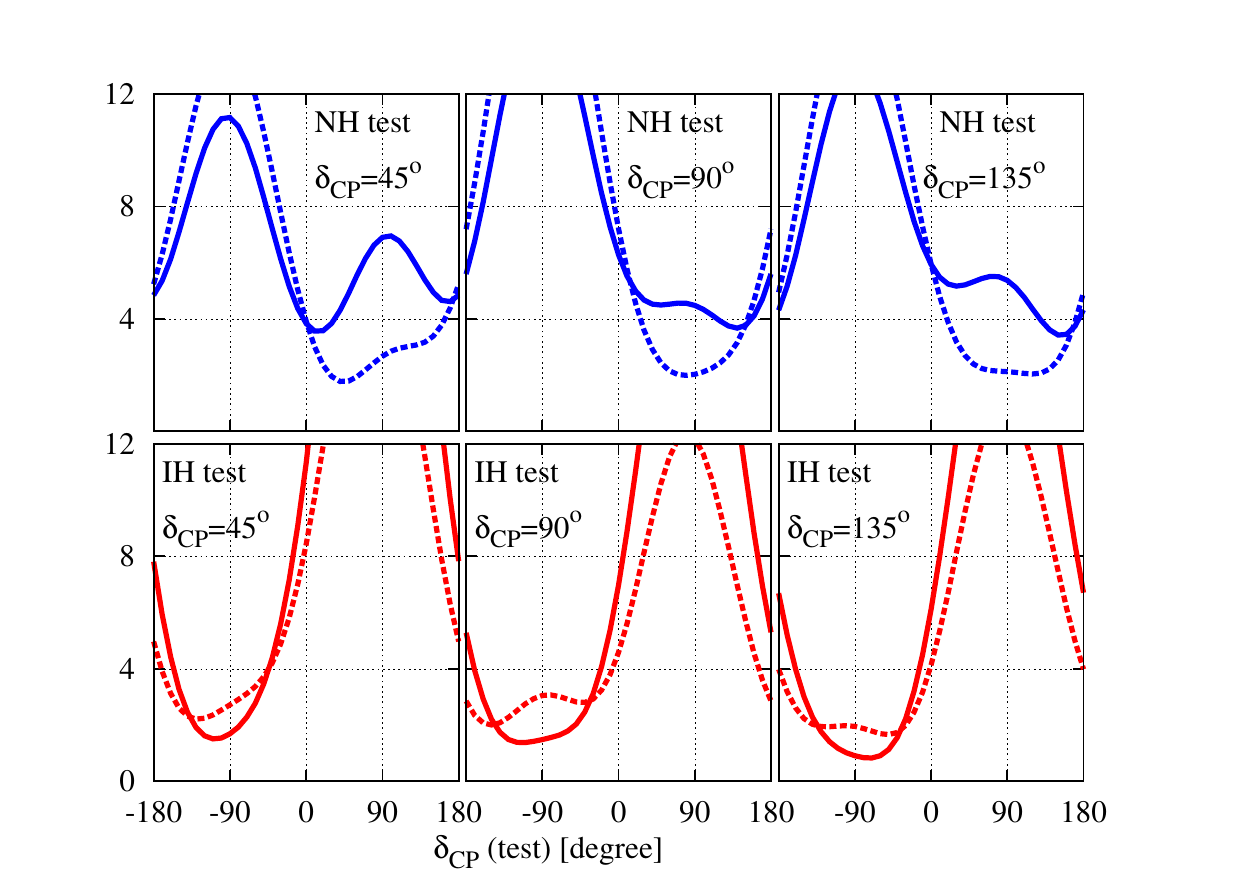}
\caption{\footnotesize{$\dchsq$ vs test $\dcp$ plot for NH true and true $\dcp$ in upper half plane. The upper
(lower) panel shows the plot for test hierarchy NH (IH). True $\dcp$ values 
are written on the panels. The solid (dashed) lines give hierarchy determination capability of \nova
as a function of test values of $\dcp$, with (without) the addition of T2K data.}}
\label{NH-unfav} 
\end{figure}

These hierarchy-$\dcp$ combinations are unfavourable for hierarchy determination 
by \nova alone. If these are the true combinations, the fit to \nova data yields two 
degenerate solutions: One with the NH and $\dcp$ in upper half plane and one with 
IH and $\dcp$ in lower half plane. The $\dchsq$ of \nova simulations for these solutions
will be the same. If we add the T2K data, which disfavours $\dcp$ in upper half plane, 
we find that the true solution of NH and $\dcp$ in the upper half plane has a rather
large $\dchsq_{\rm HR} \geq 4$ whereas the wrong hierarchy solution, IH with $\dcp$
in the lower half plane, has $\dchsq_{\rm HR} \leq 4$. This can be seen in fig.~\ref{NH-unfav}, where 
$\dchsq_{\rm HR}$ vs test $\dcp$ is plotted for test hierarchy NH (IH) in upper (lower) panel.

\subsection{IH as the true hierarchy and true $\dcp = -45^\circ,\ -90^\circ,\ -135^\circ$}
\begin{figure}[H]
\centering
\includegraphics[width=1.0\textwidth]{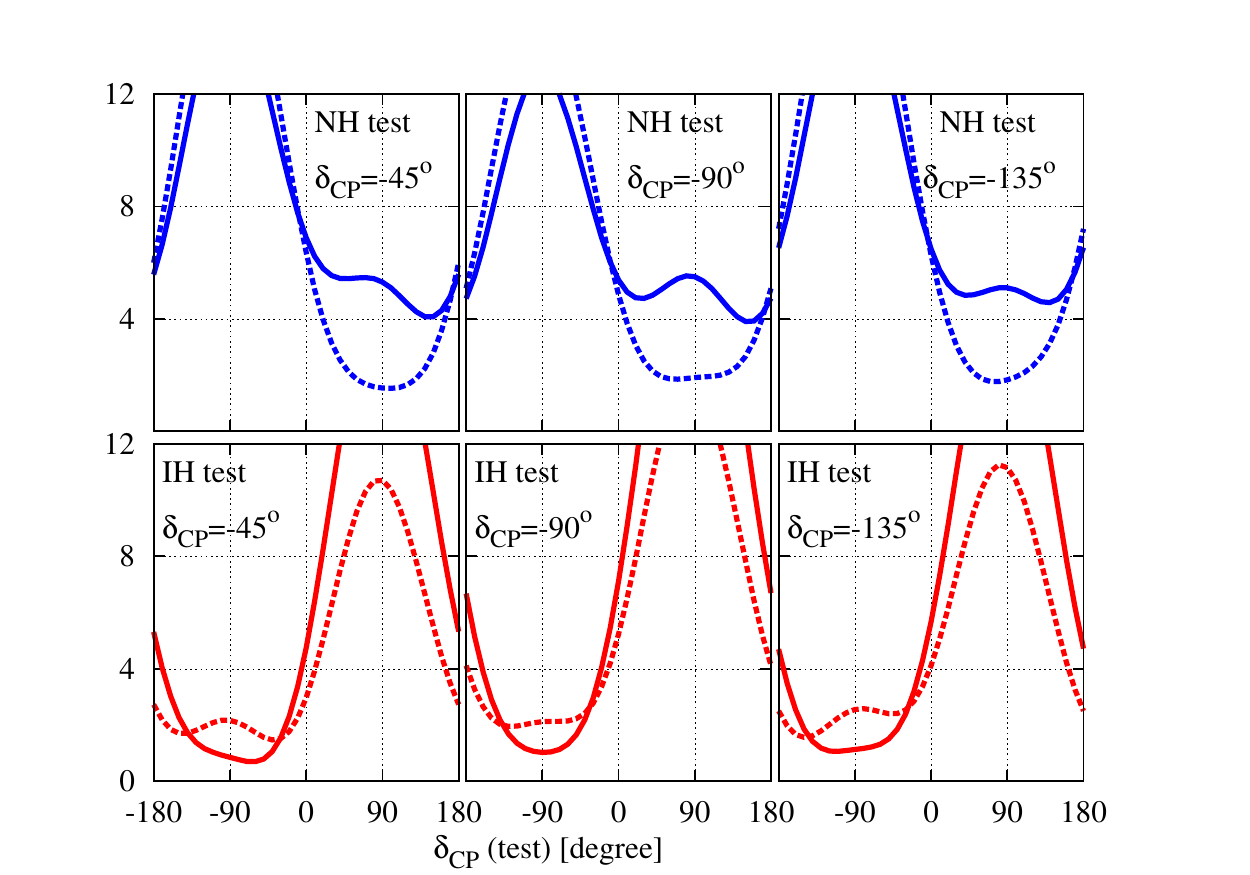}
\caption{\footnotesize{$\dchsq$ vs test $\dcp$ plot for IH true and true $\dcp$ in lower half plane. The upper
(lower) panel shows the plot for test hierarchy NH (IH). True $\dcp$ values 
are written on the panels. The solid (dashed) lines give hierarchy determination capability of \nova
as a function of test values of $\dcp$, with (without) the addition of T2K data.}}
\label{IH-unfav} 
\end{figure}

These are also unfavourable hierarchy-$\dcp$ combinations for hierarchy determination
by NO$\nu$A. For this case also, we will have degenerate solutions of NH with $\dcp$ in the 
upper half plane and IH with $\dcp$ in the lower half plane. Here the addition of T2K data
picks out the correct solution of IH with $\dcp$ in the lower half plane. The hierarchy 
determination plots are shown in fig. \ref{IH-unfav} with test hierarchy NH (IH) 
in upper (lower) panel. We see from this plot that for NH test, $\dchsq_{\rm HR}>4$ for
all test values of $\dcp$. Thus addition of T2K data with NO$\nu$a, helps to exclude
the wrong hierarchy at $2~\sigma$. 

\subsection{Hierarchy determination for true $\dcp = 0,\ 180^\circ$}
\begin{figure}[H]
\centering
\includegraphics[width=1.0\textwidth]{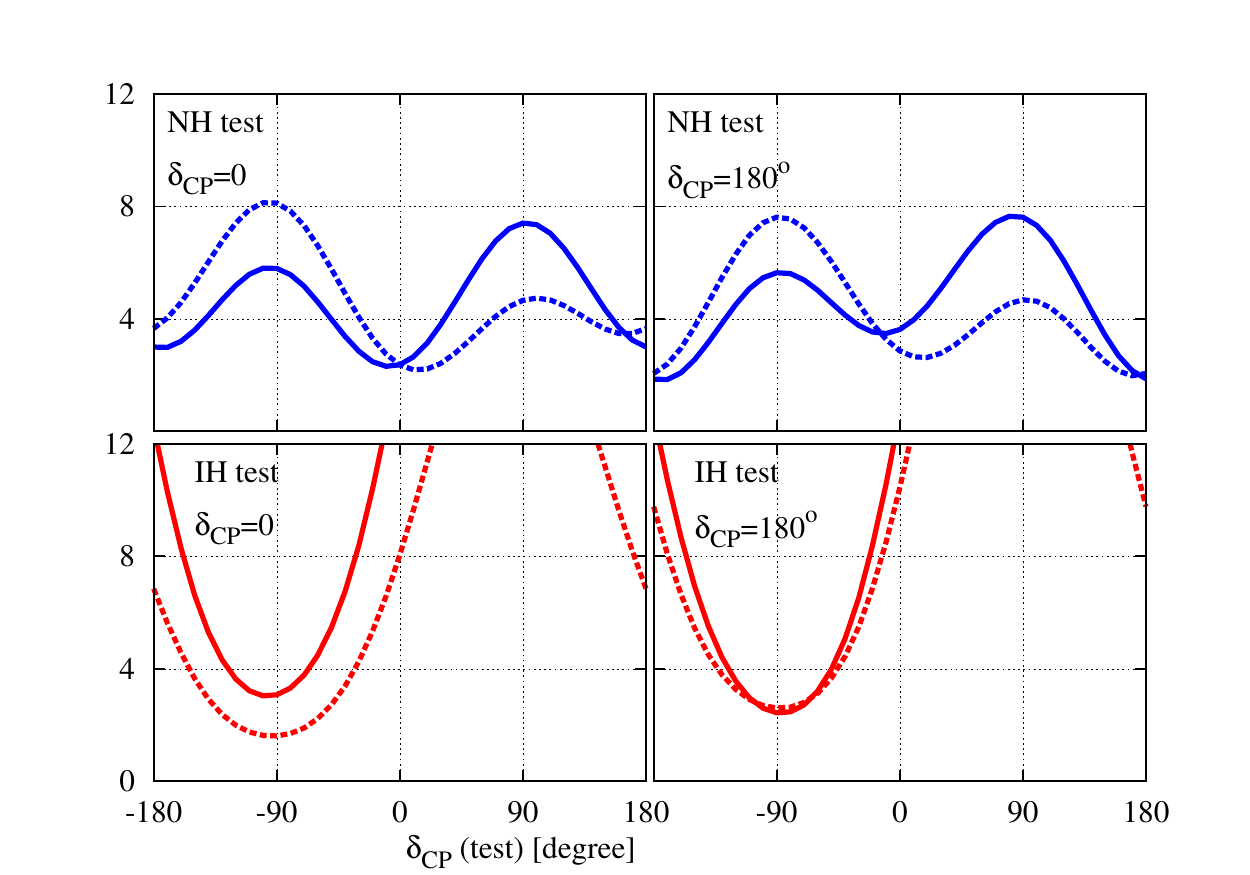}
\caption{\footnotesize{$\dchsq$ vs test $\dcp$ plot for NH true and true $\dcp$ with CP conserving values.
The left (right)
plot is for true value of $\dcp=0$ ($180^\circ$). Test hierarchy 
is NH (IH) for top (bottom) panel. The solid (dashed) lines signify 
\nova simulations combined with (without) T2K data.}}
\label{NH0}
\end{figure}

\begin{figure}[H]
\centering
\includegraphics[width=1.0\textwidth]{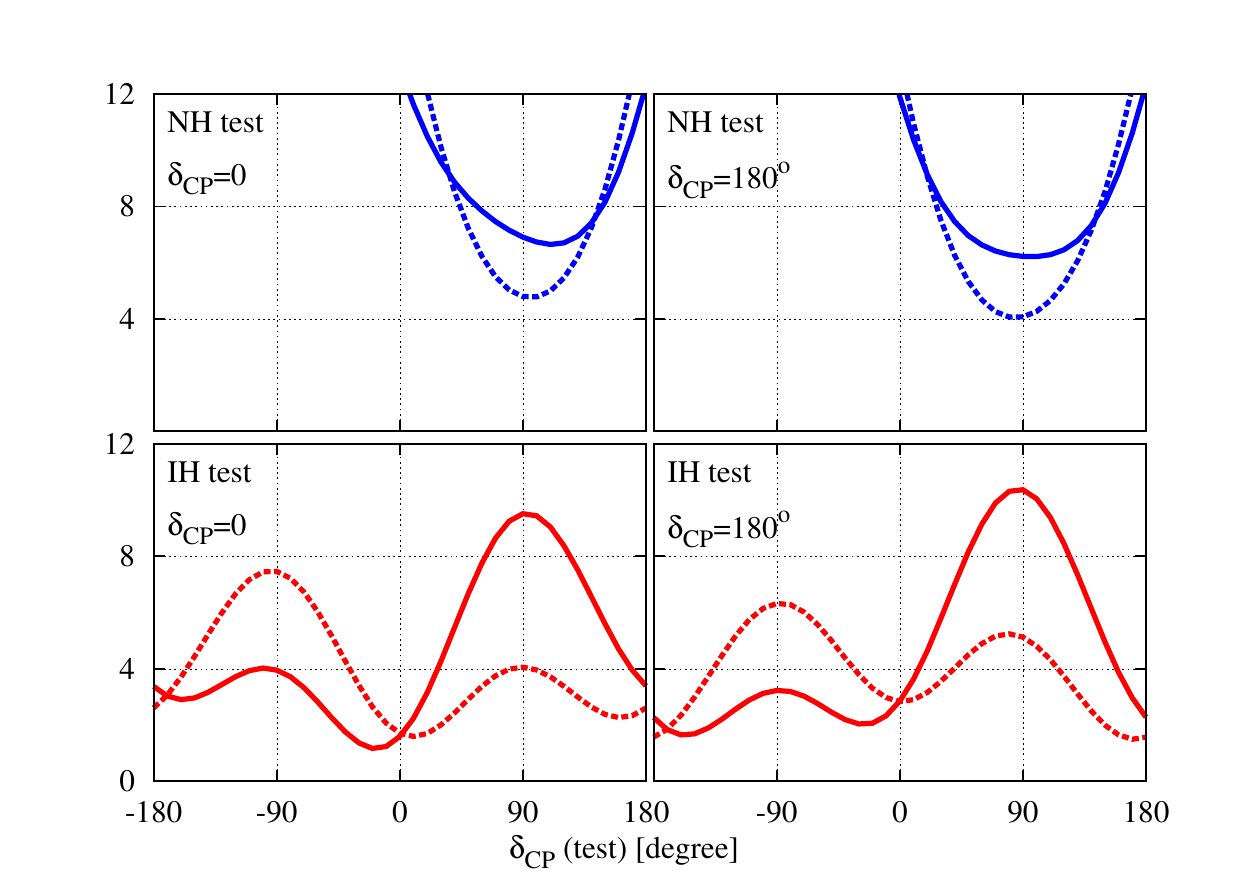}
\caption{\footnotesize{$\dchsq$ vs test $\dcp$ plot for IH true and true $\dcp$ with CP conserving values.
The left (right)
plot is for true value of $\dcp=0$ ($180^\circ$). Test hierarchy 
is NH (IH) for top (bottom) panel. The solid (dashed) lines signify 
\nova simulations combined with (without) T2K data.}}
\label{IH0}
\end{figure}

These are the CP conserving $\dcp$ values for NH true. Fig.~\ref{NH0} shows the plots with 
true $\dcp = 0$ ($180^\circ$) in left (right) panel and test hierarchy NH (IH) in upper (lower)
panel. From the figure we see that for both the CP conserving $\dcp$ values, the wrong 
hierarchy can not be excluded completely at $2\sigma$ C.L., even after the addition of
T2K data with NO$\nu$A. Thus hierarchy determination is not possible for the CP conserving 
values of $\dcp$ when NH is the true hierarchy. However, when IH is the true hierarchy,
NH can be effectively ruled out for the CP conserving $\dcp$ values, as illustrated in fig.~\ref{IH0}.

\section{Analysis of recent \nova and T2K data}
In the previous section, we studied the effect of combining the $\nu_e$
appearance data of T2K \cite{T2Kapp} with \nova simulations to estimate 
the hierarchy determination potential. Recently, T2K has published their
anti-neutrino data corresponding to an exposure of $4 \times 10^{20}$ POT 
\cite{Salzgeber:2015gua} and \nova has released the results of
their first neutrino run with an exposure of $2.7 \times 10^{20}$ POT 
\cite{novadata,Bian:2015opa}. It will be interesting to study
the neutrino parameter space allowed by these three pieces of data. 


\begin{figure}[H]
\centering
\includegraphics[width=0.8\textwidth]{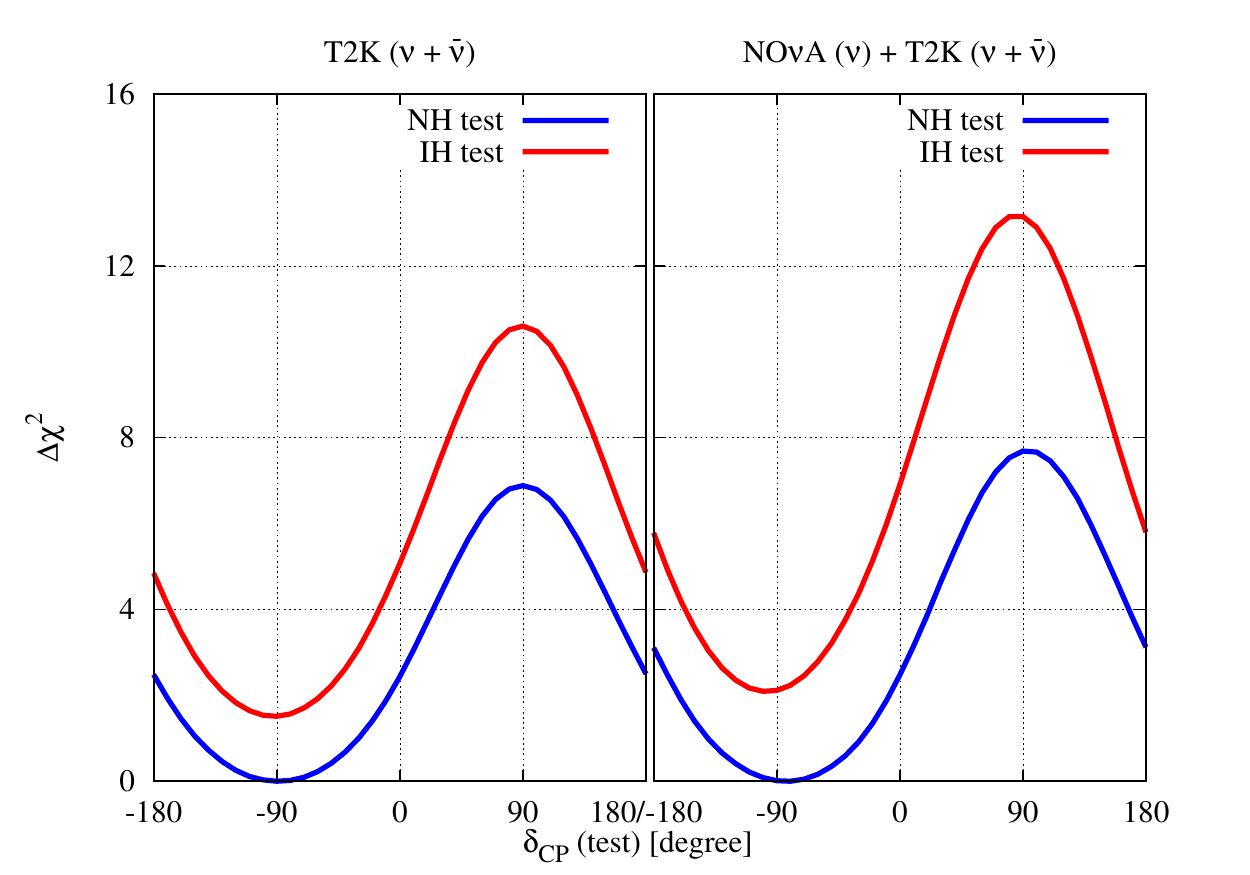}
\caption{\footnotesize{$\dchsq$ vs test $\dcp$ plot for combined analysis of T2K and \nova data.
The left (right) panel shows the analysis of T2K neutrino and anti-neutrino appearance data without (with)
the \nova neutrino appearance data.}}
\label{combined}
\end{figure}

In fig.~\ref{combined}, we have shown $\dchsq$ from 
the combined appearance data of T2K $\nu$ and $\anu$ runs
and \nova $\nu$ run, as a function of test values of $\dcp$ for both
NH and IH as test hierarchies. The results in this plot show the same features as
the results obtained from the analysis of T2K neutrino data. The best fit point
occurs for NH and $\dcp = -90^\circ$. For both the hierarchies, the lower half plane
is favoured and the upper half plane is disfavoured. In particular, a large fraction 
of the upper half plane is ruled out at $2~\sigma$ for NH and the whole of it ruled
out at $2~\sigma$ for IH. Our results match with those of ref. \cite{Palazzo:2015gja} obtained earlier.

\section{Summary and Conclusions}

In this paper we have studied influence of the present neutrino data of
T2K on the hierarchy determination ability of NO$\nu$A. This study required
combining the simulations of \nova with the data of T2K. This posed a challenge
because fluctuations are inherent in the data of T2K. We overcame this problem
by simulating the \nova data with Poissonian fluctuations. To minimize the
effect of the fluctuations, we did 100 different simulations and took the 
average. We also showed that a larger number of simulations do not change our
conclusions. 

Regarding the hierarchy determination capability of NO$\nu$A, T2K data has no effect 
if the hierarchy-$\dcp$ combinations are favourable. For such cases, \nova data determines 
the hierarchy. For the unfavourable combinations one must exercise care. For the
combination IH and $\dcp$ in lower half plane, the T2K data picks out the
correct solution between the degenerate solutions allowed by the \nova data. 
For the combination NH and $\dcp$ in the upper half plane, the T2K data favours
the wrong hierarchy-wrong $\dcp$ solution between the degenerate solutions.
If the combination of T2K and \nova data gives IH and $\dcp$ in the lower half plane
as the preffered solution, it may not be correct. It is possible that the correct
solution is NH and $\dcp$ in the upper half plane but the preference of the present
T2K neutrino appearance data for $\dcp$ in the lower half plane leads to the wrong solution.  
Hence we conclude that the present neutrino data of T2K does not help in rejecting the
wrong hierarchy, in the case of unfavourable combinations. In such a situation, 
data from an experiment such as DUNE \cite{Acciarri:2015uup} is needed to resolve the 
hierarchy-$\dcp$ degeneracy.


\bibliographystyle{apsrev}
\bibliography{referenceslist}
\end{document}